\documentclass[submission, Phys]{SciPost}

\hypersetup{
    colorlinks,
    linkcolor={red!50!black},
    citecolor={blue!50!black},
    urlcolor={blue!80!black}
}

\usepackage[bitstream-charter]{mathdesign}
\DeclareSymbolFont{usualmathcal}{OMS}{cmsy}{m}{n}
\DeclareSymbolFontAlphabet{\mathcal}{usualmathcal}

\usepackage{bm}
\usepackage{graphicx}
\usepackage{color}
\usepackage[english]{babel}
\usepackage[T1]{fontenc}
\usepackage{subeqnarray}
\newcommand{\<}{\langle}
\renewcommand{\>}{\rangle}

\newcommand{\BLUE}{\color{blue}}

\newcommand{\HH}{{\mathbb H}}
\newcommand{\SSS}{{\mathbb S}}

\begin{document}

\begin{center}{\Large \textbf{
Anomalous criticality coexists with giant cluster in the uniform forest model\\
}}\end{center}

\begin{center}
Hao Chen\textsuperscript{1$\star$},
Jes\'us Salas\textsuperscript{2,3$\dagger$} and
Youjin Deng\textsuperscript{4,5$\ddagger$}
\end{center}

\begin{center}
{\bf 1}
School of the Gifted Young, University of Science and Technology of China, \\
Hefei, Anhui 230026, P.R. China \\
{\bf 2}
Departamento de Matem\'aticas, Universidad Carlos III de Madrid, \\
Avenida de la Universidad 30, 28911 Legan\'es, Spain
\\
{\bf 3}
Grupo de Teor\'{\i}as de Campos y F\'{\i}sica Estad\'{\i}stica,
Instituto Gregorio Mill\'an Barbany,
Universidad Carlos III de Madrid. Unidad Asociada al Instituto de
Estructura de la Materia, CSIC, Serrano 123, 28006, Madrid, Spain \\
{\bf 4}
Department of Modern Physics, University of Science and Technology of China,
\\ Hefei, Anhui 230026, P.R. China\\
{\bf 5}
Hefei National Laboratory,University of Science and Technology of China,
\\ Hefei, Anhui 230088, P.R. China\\

${}^\star$ {\small \sf \BLUE chenhao123@mail.ustc.edu.cn},\quad
${}^\dagger$ {\small \sf \BLUE jsalas@math.uc3m.es},\quad
${}^\ddagger$ {\small \sf \BLUE yjdeng@ustc.edu.cn}
\end{center}

\def\today{November 10, 2023; revised: March 31, 2024}
\begin{center}
\today
\end{center}


\section*{Abstract}
{\bf
We show by extensive simulations that the whole supercritical phase
of the three-dimen\-sion\-al uniform forest model simultaneously exhibits
an infinite tree and a rich variety of critical phenomena.
Besides typical scalings like algebraically decaying correlation,
power-law distribution of cluster sizes, and divergent correlation length,
a number of anomalous behaviors emerge.
The fractal dimensions for off-giant trees take different values
when being measured by linear system size or gyration radius.
The giant-tree size displays two-length scaling fluctuations,
instead of following the central-limit theorem.
}

\vspace{10pt}
\noindent\rule{\textwidth}{1pt}
\tableofcontents\thispagestyle{fancy}
\noindent\rule{\textwidth}{1pt}
\vspace{10pt}

\section{Introduction}
\label{sec:intro}
Percolation studies connectivity
in random geometric systems~\cite{Stauffer1991,Hugo2018}.
In statistical physics, percolation has been of immense theoretical
interest, providing a simple example that undergoes
a non-trivial phase transition.
The celebrated Ising and Potts models~\cite{Wu_82}
can be described as a correlated percolation
through the exact Fortuin-Kasteleyn
transformation~\cite{Kasteleyn_69,Fortuin_72,Grimmett2006}.
Thanks to the percolation approach, it was 
established~\cite{Hugo2016a,Hugo2016b,Hugo2021,Hugo2022,Grimmett2022} that
the Ising model in three dimensions (3D)
has a sharp continuous phase transition,
and in 4D, it exhibits mean-field critical behavior,
proving the triviality of the 4D Euclidean scalar quantum field theory.
Percolation has also intensively been studied in mathematics,
including a list of variations
like $k$-core and explosive percolation, etc
\cite{Achilioptas2009,Ziff2009,Grassberger2011,Riordan2011,LiDeng2023,Ziff2023}.
Also, percolation has been applied in diverse branches of science
and industry~\cite{Ziff2014,Saberi2015}.

In the basic bond percolation, one randomly occupies each lattice
edge with probability $p$ and constructs clusters of
connected components.
Clusters are small for small $p$,
and the probability (two-point correlation) that two sites with distance
$r$ are in
the same cluster decays exponentially as $g(r) \sim \exp(-r/\xi)$,
and the correlation length $\xi$ diverges as $\xi \sim (p_c-p)^{-\nu}$
as the threshold $p_c$ is approached.
At $p_c$, the size $s$ of fractal clusters follows
a universal power-law distribution as $n(s) \sim s^{-\tau}$, and
the correlation decays algebraically as $g(r) \sim r^{2-d-\eta}$.
For the supercritical phase ($p>p_c$),
an infinite cluster of size $C_1$ occupies a nonzero fraction of
the lattice---i.e., $m \equiv C_1\, L^{-d}$ converges to
a constant as the linear size $L \to \infty$.
In percolation, $m$ plays a role as the order parameter and
behaves as $m \sim (p-p_c)^{\beta}$ as $p \downarrow p_c$.
Further, the second moment of cluster sizes,
$\chi \equiv L^{-d}\, \sum_i |C_i|^2$,
acting as the magnetic susceptibility,
diverges as $\chi \sim |p-p_c|^{-\gamma}$.
Among these critical exponents $\nu, \beta, \gamma, \eta, \tau$,
two are independent and the others can be obtained
from (hyper)scaling relationships~\cite{FFS92}.

For $p>p_c$, the infinite cluster can be even proved to be
unique--i.e., there is one and only one giant cluster.
All off-giant clusters are small with finite correlation length $\xi'$,
and the correlation $g'(r)$,
defined as the probability of two points with distance $r$ being
connected by the same cluster, with the giant cluster excluded,
vanishes exponentially.
In 2D, the supercritical and subcritical ($p<p_c$)
phases are dual to each other.
Further, the smallness of $\xi'$ predicts that
the size fluctuation of the giant cluster would obey the
central-limit theorem and follow a normal (Gaussian) distribution.

In this work, we study the percolative properties of
the supercritical phase for the (weighted) uniform forest (UF) model
in 3D~\cite{Stephen_76,Jacobsen_04,Bauerschmidt_21,Bauerschmidt_21b}.
The UF model (also called the arboreal gas)
consists of a spanning forest of trees (acyclic clusters),
in which each tree is weighted by a factor $w$ per occupied bond;
the statistical weight of any configuration ${\cal{A}}$
can be written as $\pi({\cal A}) = w^{|\cal A|} \cdot \delta_{c({\cal A}),0}$.
This is similar to that for bond percolation with probability
$p=w/(1+w)$, except the $\delta$-function constraint on
zero cyclomatic number $c({\cal A})=0$.
Further, as bond percolation, the 3D UF model undergoes
a continuous transition $w_c$ [Fig.~\ref{fig1}(a)].
It was found that $w_c = 0.43365(2)$,
$\nu=1.28(4)$, and $\beta/\nu=0.4160(6)$ \cite{Deng_07}.
Moreover, the supercritical phase also
has an infinite and unique tree \cite{Halberstam_23}, similar to 
percolation. This is confirmed in the inset of Fig.~\ref{fig1}(a),
as $m$ quickly saturating to $m_0 = 0.685(6)$ for $w=0.9>w_c$ and
the supercritical phase thus is clearly long-ranged.
Despite these analogues to bond percolation, we find that
the whole supercritical phase of the 3D UF model exhibits
the simultaneous emergence of a surprisingly rich variety of critical behaviors
and of a unique giant tree,
providing a counter example for the standard percolation scenario.

\clearpage

\section{Theoretical Insights}
The UF model is the $q\to 0$ limit
of the $q$-state Potts model~\cite{Potts_52,Baxter_book,Wu_82,Wu_84}
in the Fourtuin--Kasteleyn random-cluster
representation \cite{Kasteleyn_69,Fortuin_72,Grimmett2006}.
Particularly, this limit should be taken such that
$(e^{J}-1)/q =w$ is held fixed
($J$ is the reduced nearest-neighbor coupling of the Potts model).
Despite its simplicity, the statistical properties of the model
so far are partially understood.

A key difference between the UF model and the $q$-state Potts model is that
the UF model acquires a continuous symmetry after taking the $q\to0$ limit,
as opposed to the discrete $S_q$ symmetry of the Potts model.
Specifically, the model was mapped
onto a non-Gaussian fermionic theory with
a non-abelian continuous OSP$(1|2)$ supersymmetry
by generalizing Kirchhoff's matrix-tree theorem~\cite{Caracciolo_04}.
In addition, the authors in \cite{Caracciolo_04} showed
how to map this model in perturbation theory
to all orders in $1/w$ onto a $N$-vector model
analytically continued to $N=-1$. They concluded that in 2D,
the model is asymptotically free, implying the absence of phase 
transition for any
finite $w>0$ \cite{Jacobsen_04,Bauerschmidt_21};
the criticality occurs at $w=+\infty$ \cite{Wu_78,Baxter_book}.
The relation of the UF model and the supersphere non-linear sigma
model (NL$\sigma$M) with $\SSS^{0,2}$ was studied 
in~\cite{Jacobsen_05,Bedini_09,Caracciolo_17}.
Recently, the UF model was reinterpreted~\cite{Bauerschmidt_21}
as a NL$\sigma$M with the fermionic hyperbolic plane
$\HH^{0|2}$ as the target space.
In this field theory (see e.g., Ref.~\cite{Bauerschmidt_21c}), 
a triplet field $u_i = (\psi_i, \xi_i, z_i)$
is introduced for each site $i$ of the lattice,
with $\psi_i, \xi_i$ being Grassmann variables, 
$z_i = \sqrt{1- 2\psi_i\xi_i} = 1 - \psi_i\xi_i$ is 
a commuting variable,
and the action of this NL$\sigma$M is 
$S= (w/2) \sum_{\langle ij \rangle} (u_i - u_j)\cdot (u_i - u_j)$ (where the
dot product is defined as 
$u_i\cdot u_j = -\psi_i\xi_j - \psi_j\xi_i - z_iz_j$, so that 
$u_i\cdot u_i = -1$).
This mapping also leads to a correspondence between
the two-point correlation of the NL$\sigma$M and
the two-point connection probability
(the probability of two points being in the same cluster) in the UF model:
\begin{equation}\label{eq:corr_mapping}
     -\langle u_i\cdot u_j\rangle_{\mathbb{H}^{0|2}}=
     \mathbb{P}_{\mathrm{UF}}[i\leftrightarrow j].
\end{equation}
From the RG analysis of the $\HH^{0|2}$ model~\cite{Bauerschmidt_21b},
it was shown that there is a phase transition to the $\HH^{0|2}$  
symmetry-breaking phase for $d \ge 3$ as $w$ increases.
Furthermore, it was proven~\cite{Bauerschmidt_21b} that,
at large enough $w$ (deep in the symmetry-breaking phase),
the two-point correlation has an algebraic form due to the Goldstone
mode associated with the continuous symmetry breaking.
For the UF model, these rigorously established results translate
to the conclusion that there is a percolation transition for $d\geq 3$,
and the two-point connection probability for $d\geq 3$ at large 
$w$ has the form
\begin{equation}
g(\bm{r}) \;=\; g_0 + (c/w) \, |\bm{r}|^{2-d} + \cdots\,,
\label{eq:g}
\end{equation}
where $g_0=m^2$ comes from the giant cluster,
$c = c_0 + O(1/w)$ is a constant, and the dots stand for higher-order corrections.
An intuitive physical picture for Eq.~(\ref{eq:g}) can be formulated 
in the RG framework: 
the order parameter of the non-linear sigma model in the low-temperature phase 
has the longitudinal and transverse modes; 
the nonzero background $g_0$ comes from the longitudinal mode, 
and the power-law decaying $ \sim |\bm{r}|^{2-d}$ is attributed to the Gaussian fluctuations 
along the transverse direction. 
In addition, it was proven~\cite{Halberstam_23} that the supercritical 
phase has, in the infinite-lattice limit, a unique infinite tree for $d=3,4$.

The power-law correlation in Eq.~(\ref{eq:g}) hints that the model also exhibits
criticality in the supercritical phase.
However, the connection to the $\mathbb{H}^{0|2}$ model so far is unable to 
provide more precise
descriptions of the critical geometric properties of the supercritical phase.
The percolative properties such as the scaling behaviors of the off-giant
clusters, the cluster-size distribution, and the probability distribution 
of the giant cluster remain elusive in the supercritical phase.
Furthermore, for a finite system of side length $L$,
it is not clear how the finite-size critical properties would behave. 
In particular, since the finite-size scaling (FSS) theory for 
a second-order phase transition is based on the assumption that  
the diverging correlation length---the only length scale near a 
critical point---is smoothed out to be of order $O(L)$, 
it is not clear whether the standard FSS theory would 
apply in the supercritical phase. 
The goal of this paper is not limited to the numerical verification of 
the existing theoretical predictions, but also to provide novel insights 
into these unknown questions from extensive numerical simulations.

\section{Results}

\begin{figure}[t]
\centering
\includegraphics[width=0.85\columnwidth]{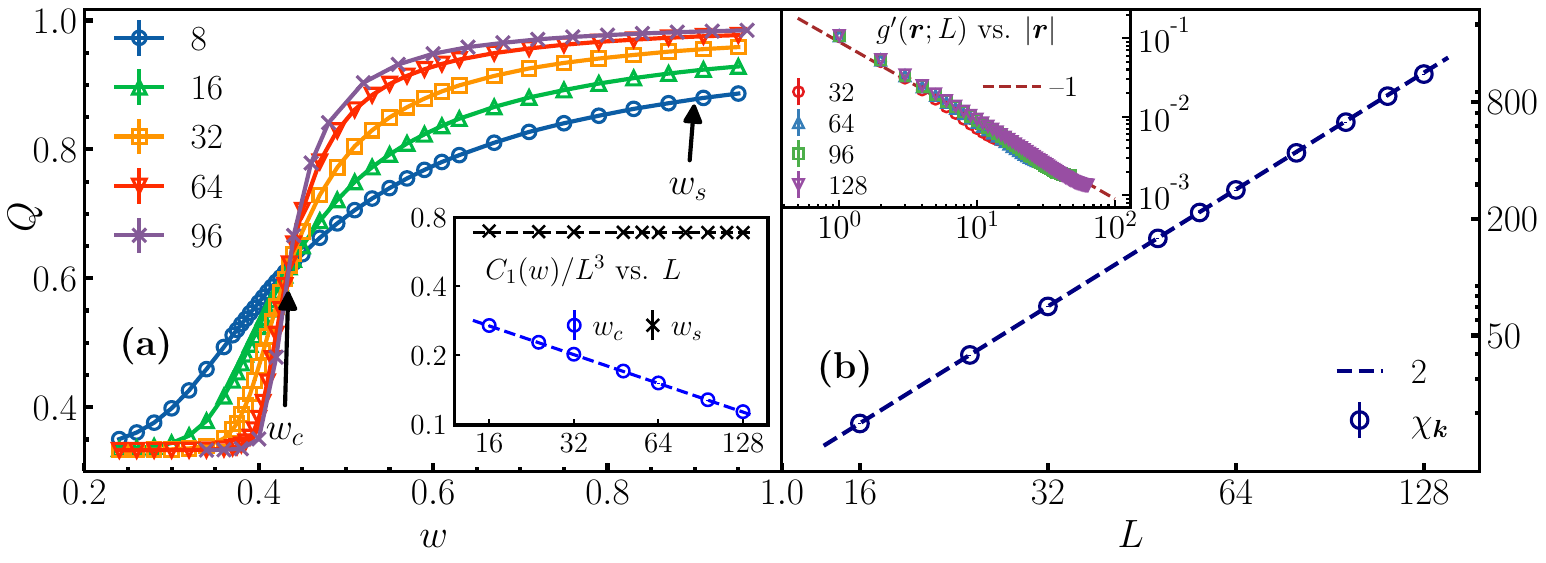}
\vspace*{-2mm}
\caption{
Simultaneous emergence of critical behaviors and a giant cluster.
(a) The percolation transition and the giant tree of size $C_1$
in the supercritical phase.
The approximately common intersection of the ratios $Q$ for different sizes $L$
indicate the threshold at $w_c \approx 0.43365$.
The inset shows that, while the order parameter $m (w_c) \equiv C_1\, L^{-3}$
algebraically vanishes, it quickly saturates to a constant,
which is $ m_0 = 0.685(6)$ for $w_s=0.9$.
(b) Emergent critical behaviors at $w=w_s$ demonstrated by
the power-law divergence of the Fourier-transformed
susceptibility $\chi_{\bm{k}} \sim L^{1.99(2)}=L^2$,
and the algebraic decaying of the off-giant correlation
$g'(\bm{r}) \sim |\bm{r}|^{2-d} $.
}
\label{fig1}
\end{figure}

\begin{figure}[!t]
     \centering
     \includegraphics[width=0.90\columnwidth]{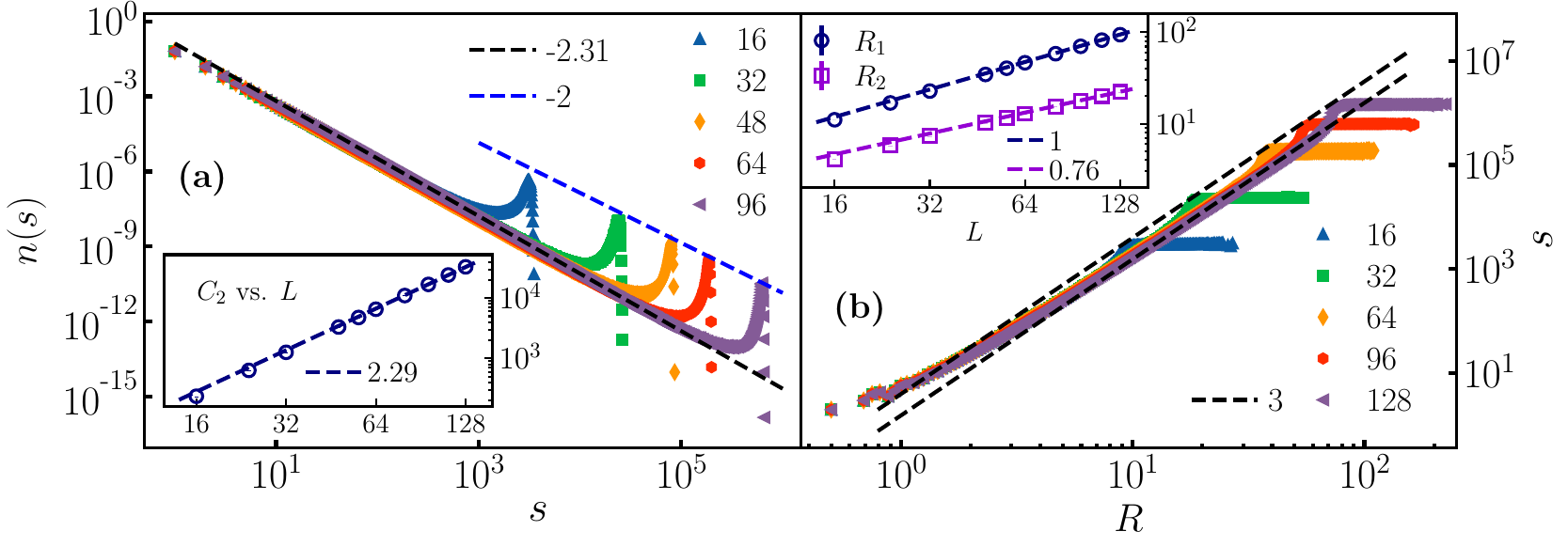}
     \vspace*{-2mm}
     \caption{%
     Fractal structures and two-length scales at $w=w_s$.
     (a) Scaling behavior of the cluster-size distribution \eqref{eq:ns}.
     The inset displays $C_2 \sim L^{d_{C_2}}$ and
     with a line of slope $d_{C_2} = 2.29$.
     (b) Power-law dependence of cluster size $s$ on the gyration radius $R$.
     The curves seem to collapse around two parallel lines of slope $d_{f}=3$.
     The inset shows the radii of the two largest clusters,
     of which the fits give exponents $\kappa_1=0.999(4)$, and
     $\kappa_2=0.76(2)$.
     }
\label{fig2}
\end{figure}

In the supercritical phase, the off-giant clusters are fractal
and display critical scaling
behaviors that are generally expected from percolation theory at criticality.
(i) As shown in Fig.~\ref{fig1}(b),
the Fourier-transformed susceptibility $\chi_{\bm{k}}$
(the definition will be given later) diverges as
$\chi_{\bm{k}} \sim L^{\gamma/\nu} = L^{1.99(2)}$.
This result is in good agreement with $\chi_{\bm{k}} \sim L^2$,
following  the standard finite-size scaling (FSS) \emph{Ansatz}.
The off-giant correlation $g'({\bm r})$ algebraically decays
as $g'(\bm{r}) \sim |\bm{r}|^{2-d-\eta}$ with $\eta=0$ \cite{Bauerschmidt_21b}.
Indeed, the scaling relation $2-\eta = \gamma/\nu$ is satisfied.
(ii) The size distribution of the clusters $n(s,L)$ has two terms: one
accounts for the contribution of the off-giant trees, while the other takes
care of the giant one. The former term contains a power law $s^{-\tau}$ times
the size distribution of the off-giant clusters $\tilde{n}'(s,L)$, which is
governed by the second-largest cluster of size $C_2 \sim L^{d_{C_2}}$ with
$d_{C_2}=2.29(2)$. The latter term has a prefactor $L^{-d}$, and it is
governed by the size of the largest cluster
$C_1 \sim L^{d_{C_1}}=L^{3.000(2)} = L^d$. Putting all together, we have
[see Fig.~\ref{fig2}(a)]:
\begin{equation}
n(s,L) \;=\; s^{-\tau} \, \tilde{n}'(s\, L^{-d_{C_2}})+ L^{-d}\, n_1(s, L)
\,.
\label{eq:ns}
\end{equation}
The Fisher exponent is $ \tau \equiv \tau_2 =  2.31(2)$,
and the distribution peak,
arising from $C_1$, defines an effective Fisher exponent
$\tau_1 = 2$. The hyperscaling relations
$\tau_i = 1+d/d_{C_i}$ for $i=1,2$ are satisfied.

In addition, the supercritical phase  exhibits a variety
of unusual critical behaviors.
(iii) The standard FSS theory predicts that
the critical correlation length is $\xi \sim O(L)$,
and, indeed, the gyration radius $R_1$ of
the largest tree scales as $R_1 \sim L$ for $w\geq w_c$.
However, for $w>w_c$, the off-giant correlation length,
characterized by the gyration radius $R_2$ of the
second-largest cluster,
diverges sublinearly as $R_2 \sim L^{\kappa_2}$
with $\kappa_2=0.76(2)$ [see the inset of Fig.~\ref{fig2}(b)].
This indicates that the supercritical phase has
two length scales---i.e., $L$ and $ L^{0.76} \ll L$.
Typically, the size $s$ of a fractal object depends on
its gyration radius $R$ as $s \sim R^{d_f}$,
and the generic fractal dimension $d_f$ of the giant cluster
coincides with the finite-size fractal dimension $d_{C_1}$
since $R_1\sim L$.
(iv) For the off-giant clusters at $w>w_c$,
however, the generic and finite-size fractal dimensions
take different values---e.g., $ d_{C_2} \neq d_{f_2}$
for the second-largest cluster.
Instead, one has $d_{f_2}=d_{C_2}/\kappa_2$,
since $C_2 \sim L^{d_{C_2}} \sim R_2^{d_{f_2}}$
and $R_2 \sim L^{\kappa_2}$.
We obtain $ d_{f_2}= 3.01(8) \approx 3$,
which agrees well with $d$ and $d_{C_1}$;
this is further demonstrated for all the off-giant clusters
in Fig.~\ref{fig2}(b). Thus, the off-giant clusters
share the same generic fractal structure as the giant one.

\begin{figure}[!t]
     \centering
     \includegraphics[width=0.85\columnwidth]{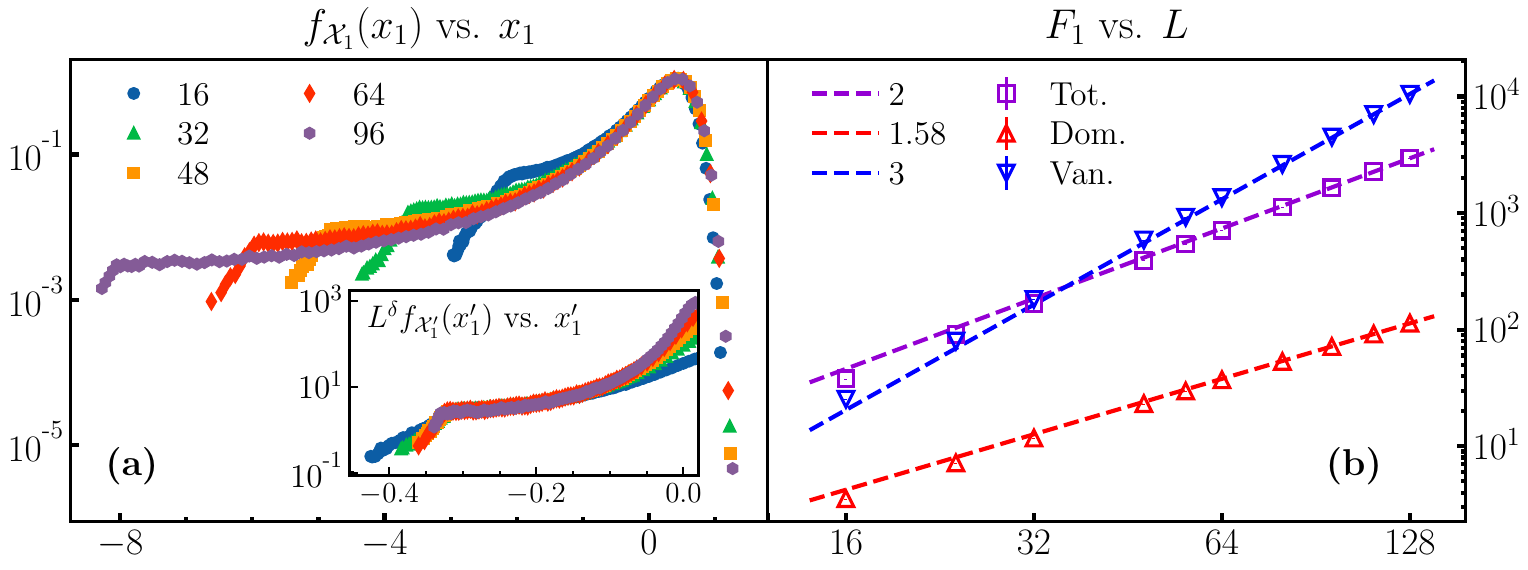}
     \vspace*{-2mm}
     \caption{Two-length-scale fluctuation of $\mathcal{C}_1$ at $w=w_s$.
     (a) PDF $ f_{\mathcal{X}_1}(x_1,L)$
     of the rescaled random variable
     $\mathcal{X}_1\equiv (\mathcal{C}_1 - \< \mathcal{C}_1 \>)\, L^{-d_{C_2}}$
     with $d_{C_2} =2.29$.
     A good collapse of data is observed in the peak region.
     The inset shows the PDF $ L^\delta\, f_{\mathcal{X}'_1}(x_1')$
      of the random variable
     $\mathcal{X}_1' \equiv (\mathcal{C}_1 - \< \mathcal{C}_1 \>)\, L^{-3}$
     with $\delta=0.77$, showing an approximate collapse in the shoulder region.
     (b) Normalized variance $F_1 =\mathrm{Var}(\mathcal{C}_1) \, L^{-3}$ in
     different sectors.
     The fits give exponents $d_{F_1} (\text{total})=2.03(6)$,
     $d_{F_1} (\text{dominant})=1.58(2)$, and
     $d_{F_1} (\text{vanishing})=2.96(4)$.
}
\label{fig3}
\end{figure}

The giant cluster, occupying about $70\%$ of the lattice
for $w = 0.9$, exhibits interesting critical behaviors.
In a traditional supercritical phase with finite correlation length,
the central limit predicts that the giant-cluster size should
follow a normal (Gaussian) distribution and the normalized fluctuation
$F_1 \equiv {\rm Var}({\cal C}_1)\, L^{-d}$ should converge to some constant.
(v) As in Fig.~\ref{fig3}(b), however, we find $F_1 \sim L^{d_{F_1}}$
with $ d_{F_1} = 2.03(6)$.
This is counter intuitive because one would naively expect
the central-limit theorem to apply
because the ratio $\xi'/L \sim L^{\kappa_2-1} \to 0$ as $L\to\infty$
(here $\xi'$ is the off-giant correlation length).

We then consider the probability density function (PDF)
$f_{\mathcal{C}_1}(C_1,L)$ of the random size $\mathcal{C}_1$ of the giant
cluster in a lattice of linear size $L$.
In the standard FSS theory,  by rescaling
$ \mathcal{X}=(\mathcal{C}_1 -\< \mathcal{C}_1\>)\, L^{-d_{C_1}}$,
and transforming $f_\mathcal{X}(x) \, dx = f_{\mathcal{C}_1}(C_1,L)\, dC_1$,
one can obtain a universal and $L$-independent function $f_\mathcal{X}(x)$.
(vi) In the supercritical phase of the UF model,
however, we find that the $ f_{\mathcal{C}_1}(C_1,L)$ data
for different values of $L$ cannot be collapsed onto a unique
curve by any single exponent like $d_{C_1}=3$.
Thus, we consider the probability $ f_{\mathcal{X}_1}(x_1,L) \, dx_1$
for the rescaled random deviation
$\mathcal{X}_1 \equiv (\mathcal{C}_1-\<\mathcal{C}_1\>)\, L^{-d_{C_2}}$
with $d_{C_2}=2.29$.
Then, the $ f_{\mathcal{X}_1}(x_1,L)$ data approximately collapses
 well near $x_1= 0$ and for $x_1 >0$ [Fig.~\ref{fig3}(a)].
Nevertheless,
$ f_{\mathcal{X}_1}(x_1,L)$ has a  wide-range shoulder
for $x_1 \ll 0$, for which an approximate data collapse
can be achieved by
$ L^\delta \, f_{\mathcal{X}_1'}(x_1')\,  dx_1' \equiv
f_{\mathcal{X}_1}(x_1,L) \, dx_1$ with $\mathcal{X}_1' \equiv
(\mathcal{C}_1-\<\mathcal{C}_1\>)\, L^{-3}$ and $\delta=0.77$.
This means that the whole configuration space
is roughly partitioned into two sectors: one takes up a finite
configuration-space volume
while the other vanishes asymptotically as $L^{-\delta}$.
In the dominant sector, the critical fluctuation of ${\cal C}_1$
is governed by $d_{C_2}$ for off-giant clusters.
In the vanishing sector, the variance ${\rm Var}({\cal C}_1)$ is
$\sim O(L^{2d})$. Note that this exponent takes the largest
possible value.
We further sample ${\rm Var}({\cal C}_1)$ conditioned on
$ {\cal C}_1 -C_1 \geq 0$
for the dominant (dom) sector,
and $L^{-d}({\cal C}_1 -C_1) \leq a $ for the vanishing (van) sector, where
we take $a=-0.1$.
We obtain $d_{F_1} (\mathrm{van}) =2.96(4)$
and $d_{F_1} (\mathrm{dom}) = 1.58(2)$ [see Fig.~\ref{fig3}(b)],
the latter of which gives $d_{C_2}=2.29(1)$
from relation $2 d_{C_2}-3 = d_{F_1}$.
Note that $d_{F_1}({\rm total})=2.03(6)$ for the total configuration
space is distinct from $d_{F_1}(\mathrm{dom})$ or $d_{F_1} (\mathrm{van})$,
indicating that the crossover regime also plays an important role.

\section{Algorithms and observables}
We use the Sweeny algorithm~\cite{Sweeny_83}
and simulate the UF model on the simple-cubic lattice
with periodic boundary conditions, for $8 \le L \le 128$.
At every step, one randomly picks up an edge $e_{ij}$
between sites $i$ and $j$.
If $e_{ij}$ is occupied, the bond is removed with probability $\min(1,1/w)$.
If $e_{ij}$ is empty, it is occupied with  probability $\min(1,w)$ if
$i$ and $j$ belong to different trees,
and, otherwise, it is left unoccupied since
a bond on $e_{ij}$ would generate a cycle.
The nontrivial operation is to detect the connectivity of $i$ and $j$ in a
dynamical setting. Using the link-cut tree data structure~\cite{Sleator_83},
the connectivity query can be efficiently implemented in
$O(\log L)$ amortized time. Note that no critical slowing down occurs in 3D
\cite{Sweeny_83,Deng_07,Deng_07b}.

For a random configuration, we denote the forest of trees as $\{ {\cal C}_k \}$,
and specifically leave $\mathcal{C}_1$ and $\mathcal{C}_2$ for the largest
and second-largest clusters, respectively.
For a tree $\mathcal{C}_k$, an arbitrary site is chosen as the origin,
and the ``unwrapped'' coordinate $\bm{x}_k^i$ of each site $i$
is obtained by growing the tree from the origin.
 This coordinate $\bm{x}_k^i$ is well defined, since the path
connecting any two sites in  a tree is unique.
The mass-center coordinate
$\overline{\bm{x}}_k = (1/{\cal C}_k) \sum_{i\in \mathcal{C}_k} \bm{x}_k^i$,
and the squared gyration radius
${\cal R}_k^2 =  (1/{\cal C}_k) \sum_{i\in \mathcal{C}_k}
                 (\bm{x}^i_k -\overline{\bm{x}}_k)^2$
are calculated.

By detecting the connectivity between sites $i$ and $j$ over configurations,
we measure the two-point correlation $g({\bm r}=\bm{r}_i - \bm{r}_j)$,
as well as the off-giant correlation $g'({\bm r})$,
where $\bm{r}_i$ is the standard Euclidean coordinate of site $i$.
For simplicity, we choose $\bm{r} = (r, 0, 0)$ along the $x$-axis.
Moreover, to study the algebraic decaying behavior of the correlation,
an auxiliary Ising spin $s_i\in  \{\pm 1\}$ is introduced for
every site $i$: Independently for each tree, we assign the same value $s_i=1$
or $-1$ with equal probability to all the sites in the tree.
By definition, $g(\bm{r}_i - \bm{r}_j)= \< s_i s_j \>$.
The magnetization ${\cal M} = \sum_m s_m$
and its Fourier transform  ${\cal M}(\bm{k})= \sum_m s_m
\exp (i \bm{k} \cdot \bm{r}_m)$ are sampled,
where the summation is over the  whole lattice.
The smallest nonzero momenta in the $x$ direction,
$\bm{k}=(2\pi/L,0,0)$ is used for simplicity.

Statistical average and probability distribution are then taken over
the configurations generated in simulations---e.g., $C_k \equiv
\< {\cal C}_k \>$ and $R_k \equiv \< {\cal R}_k \>$ for $k=1,2$.
Also, we define the normalized fluctuation
$ F_1 \equiv \mathrm{Var}(\mathcal{C}_1)\, L^{-3}$,
the susceptibility $\chi \equiv \< {\cal M}^2\>\, L^{-3}$,
the dimensionless ratio $Q \equiv \< {\cal M}^2\>^2/\< {\cal M}^4\>$,
and the Fourier-transformed susceptibity  $ \chi_{\bm k} \equiv
\< {\cal M}({\bm k}) {\cal M}(-{\bm k})\> \, L^{-3}$.

\section{Fits}
As a powerful quantity for locating a continuous phase transition,
the crossings of the $Q(w)$ curves for different sizes $L$
[Fig.~\ref{fig1}(a)] clearly support the previously determined
percolation threshold $w_c = 0.43365(2)$ \cite{Deng_07}.
We carry out extensive simulation at $w_c$
and $w_s=0.9$, deeply in the supercritical phase.
The critical behaviors, shown in Figs.~\ref{fig1}--\ref{fig3},
have been qualitatively presented in~{\em Results}.

\begin{table}[!b]
\centering
\setlength\tabcolsep{3.0pt}
\begin{tabular}{|r|ll|ll|ll|}
\hline
     & \multicolumn{1}{c}{$d_{C_1}$} &
       \multicolumn{1}{c|}{$\kappa_1$} &
       \multicolumn{1}{c}{$d_{C_2}$} &
       \multicolumn{1}{c|}{$\kappa_2$} &
       \multicolumn{1}{c}{$d_{F_1}$} &
       \multicolumn{1}{c|}{$d_{\chi_{\bm k}}$} \\ \hline
tot. & 3.000(2) & 0.999(4) & 2.29(2) & 0.76(2)  & 2.03(6)  & 1.99(2)   \\
dom. & 3.002(3)  & 1.001(5) & 2.28(2) & 0.78(2)  & 1.58(2)  & 1.63(3)   \\
van. & 2.997(4)  & 0.997(6) & 3.00(2) & 1.01(2)  & 2.96(4)  & 2.83(5)   \\
\hline
\end{tabular}
\caption{Estimated critical exponents for the supercritical phase
at $w_s=0.9$, for the total (tot.) configuration space, and for the
dominant (dom.) and vanishing (van.) sectors.
}
\label{tab:exponent}
\end{table}
We perform the least-squares fits to a power-law
\emph{Ansatz} for any observable $\mathcal{O}(L)$
\begin{equation}
\mathcal{O}(L) \;=\; L^{d_\mathcal{O}}\, \left( a_0 + a_1 L^{-\omega_1} +
   a_2 L^{-\omega_2} \right) + b_0 \,.
\label{def_ansatz}
\end{equation}
In most cases, we set $\omega_1=1$ and $\omega_2=2$,
and $b_0=0$ is fixed for observables that vanish for $L \to \infty$.
As a precaution against FSS corrections not included in the \emph{Ansatz}
\eqref{def_ansatz},
we have performed each fit by allowing only data with $L\ge L_\text{min}$. By
studying how the estimates of the parameters, as well as the $\chi^2$
per degree of freedom, vary as a function of $L_\text{min}$, we
determine our final estimates and their error bars.
In particular, we consider the sizes and radii of
the largest and second-largest clusters,
the normalized fluctuation $F_1$,
and the Fourier-transformed susceptibility $\chi_{\bm{k}}$,
which scale as ($j=1,2$)
\begin{eqnarray}
C_j & \sim & L^{d_{C_j}} , \;\; R_j \sim L^{\kappa_j} ,
\;\; F_1 \sim L^{d_{F_1}} , \;\;
\chi_{\bm{k}} \sim L^{d_{\chi_{\bm k}}} \,. \quad
\end{eqnarray}

At the critical value $w_c$, the scaling behaviors follow the standard FSS 
theory, which predicts $\kappa_1=\kappa_2=1$, $d_{C_1}=d_{C_2}=d_f$
($d_f$ is the generic fractal dimension), and $d_{F_1}=2 d_f-d$.
There is only one non-trivial exponent $d_{C_1}$,
which is determined to be $d_{C_1}=2.5840(6)$.

In the supercritical phase with $w=w_s$,
the final estimates are given in Table~\ref{tab:exponent}.
As expected, the largest cluster, occupying a finite fraction
of the lattice, has trivial exponents
$d_{C_1}=3.000(2)=3$ and $\kappa_1 =0.999(4)=1$.
The effective Fisher exponent $\tau_1$,
governing the decreasing of the distribution peak in Fig.~\ref{fig2}(a),
is also trivial $\tau_1 = 1+d/d_{C_1}=2.000(1)$.

The finite-size fractal exponent of the second-largest cluster is
$d_{C_2}=2.29(2)$, which has not yet been reported to our knowledge.
This gives the Fisher exponent $\tau_2=1+d/d_{C_2}=2.31(2)$ in 
Eq.~\eqref{eq:ns}.
The gyration radius scales sublinearly versus $L$
with exponent $\kappa_2=0.76(2)$,
unexpected from the standard FSS theory.
The generic fractal dimension, $C_2 \sim R_2^{d_{f_2}}$,
is calculated as $d_{f_2}=d_{C_2}/\kappa_2 = 3.01(8)=3$.
Surprisingly, $d_{f_2}$ is just the spatial dimension,
and Fig.~\ref{fig2}(b) further gives
$d_{f_2}=3$ for all the off-giant clusters.

By definition, the Fourier-transformed magnetic 
susceptibility is $ \chi({\bm k})=
L^{-3}\, \sum_{m,n} \< s_m s_n \>$ $\exp(i \bm{k}\cdot(\bm{r}_m-\bm{r}_n))$,
where the contribution from the background term $g_0$ in Eq.~\eqref{eq:g}
is eliminated.
Thus, $\chi({\bm k}) \sim L^2$ is expected, and this is strongly
supported by the estimated exponent $d_{\chi_{\bm k}}=1.99(2)$.
Interestingly, the normalized fluctuation of the largest tree
is governed by exponent $d_{F_1}=2.03(6)=2$.

Despite the simplicity of scalings like $C_1 \sim L^3$
and $F_1 \sim L^2$, the distribution of $\mathcal{C}_1$ is
sophisticated [Fig.~\ref{fig3}(a)].
Thus, we perform separate least-squares fits
for the dominant and the vanishing sectors,
respectively conditioned on ${\cal C}_1-C_1 \geq 0$ and
$ {\cal C}_1 -C_1 \leq -0.1\, L^{3}$.
The results in Table~\ref{tab:exponent} suggest that
the two sectors have dramatically different scaling behaviors.
Particularly, in the vanishing sector, we find that
both the largest and the second-largest clusters
have $d_{C_j}=3$ and $\kappa_j=1$  for $j=1,2$.
Further, $d_{\chi_{\bm k}}=2.83(5)$ suggests
that the $r$-dependent decaying of correlation $g(r)|_\mathrm{van}$
is extremely slow (i.e., $g(r)|_\mathrm{van} \sim r^{-0.17}$).
This behavior, together with $d_{C_j}=3$,
gives a strong hint for a logarithmic decay:
$g(r)|_\mathrm{van} \sim 1/\log(r)$.

\section{Conclusion}
While undergoing a typical continuous percolation transition,
the 3D UF model exhibits a variety of critical behaviors
in the supercritical phase.
The simultaneous existence of anomalous criticality
and of a unique giant cluster is unexpected from the standard
percolation theory.
The critical scaling behaviors not only
appear in the off-giant clusters,
but also in the fluctuation of the giant tree.
Unlike a conventional critical point,
the whole configuration space can be approximately divided into
two configuration sectors of distinct critical exponents.
Further, the overall scaling behaviors arise from some
delicate interplay of the two sectors and of the crossover
regime in between.

Some insight can be borrowed from the fermionic field theory,
but a complete and deep understanding is still needed for
this extremely rich critical behavior.
As a return, we believe that our work may also bring some
insight for critical phenomena in the Potts model
and the nonlinear-sigma model,
which are two important classes of systems in statistical mechanics
and condensed-matter physics.
For instance, for the XY model with long-range interaction,
which was recently found~\cite{Giachetti_21,Giachetti_22}
to exhibit critical behaviors in the low-temperature phase,
it would be desired to study such critical behaviors
from percolation perspective.

Some open quesions arise.
For instance, what is the upper spatial dimensionality $d_u$
for the supercritical-phase criticality for the UF model?
It is known that the zero-temperature UF model
(the uniform tree model) has $d_u=d_c=4$
and, at criticality, it has $d_u=d_p=6$.
Thus, $d_c=4$ or $d_p=6$ can equally serve as a candidate of $d_u$
for the supercritical UF model.
From recent studies for the Fortuin-Kasteleyn representation
of the Ising model~\cite{Fang_22,Fang_23},
we may have the third scenario that
the  low-temperature UF model has simultaneously
two upper dimensions at both $d_c=4$ and $d_p=6$.

\section*{Acknowledgements}
  This work was initiated by private communications with
  Tyler Helmuth, Roland Bauerschmidt, and Nicholas Crawford,
  to whom we are indebted.


\paragraph{Funding information}
  H.C and Y.D. have been supported by the National Natural Science
  Foundation of China (under Grant No.~12275263),
  the Innovation Program for Quantum Science and Technology
  (under grant No.~2021ZD0301900), Natural Science Foundation
  of Fujian province of China (under Grant No.~2023J02032).
  J.S. was partially supported by Grant No.~PID2020-116567GB-C22
  AEI/10.13039/501100011033.




\nolinenumbers

\end{document}